\documentclass[floatfix,twocolumn,prd,showpacs,preprintnumbers,amsmath,nofootinbib,amssymb,superscriptaddress]{revtex4-2}

\usepackage[utf8]{inputenc}
\usepackage{graphicx}

\newcommand{\stackss}[2]{{\begin{subarray}{l}{\textrm{\tiny #1}}\\[-0.3ex] \textrm{\tiny #2}\end{subarray}}} 

\usepackage{hyperref}
\hypersetup{colorlinks,citecolor=blue,urlcolor=blue,linkcolor=black}
\usepackage{float}
\usepackage[nameinlink,capitalize]{cleveref}
\usepackage{tcolorbox}

\crefname{section}{Sec.}{Secs.}
\crefname{equation}{Eq.}{Eqs.}
\def\ccite#1{Ref.~\cite{#1}} 
\def\ccites#1{Refs.~\cite{#1}} 

\begin{document}


\title{
    Lattice \texorpdfstring{$\boldsymbol{B}$}{B}-field correlators for heavy quarks
}

\author{Luis Altenkort$^*$}
\affiliation{Fakult\"at f\"ur Physik, Universit\"at Bielefeld, D-33615 Bielefeld, Germany }
\thanks{altenkort@physik.uni-bielefeld.de}

\author{David de la Cruz}
\affiliation{Institut f\"ur Kernphysik, Technische Universit\"at Darmstadt\\
    Schlossgartenstra{\ss}e 2, D-64289 Darmstadt, Germany }

\author{Olaf Kaczmarek}
\affiliation{Fakult\"at f\"ur Physik, Universit\"at Bielefeld, D-33615 Bielefeld, Germany }

\author{Guy D.~Moore}
\affiliation{Institut f\"ur Kernphysik, Technische Universit\"at Darmstadt\\
    Schlossgartenstra{\ss}e 2, D-64289 Darmstadt, Germany }

\author{Hai-Tao Shu}
\affiliation{Physics Department, Brookhaven National Laboratory, Upton, New York, USA}


\def\d{\mathrm{d}}

\def\tauf{\tau_\mathrm{F}}
\def\MSbar{\overline{\text{MS}}}
\def\MSBAR{\MSbar}
\def\gsqms{g^2_{_{\overline{\mathrm{MS}}}}} %
\def\alphas{\alpha_{\mathrm{s}}} %
\def\Tr{\,\mathrm{Tr}\:}
\def\OO{\mathcal{O}}
\def\gammaE{\gamma_{_{\mathrm{E}}}} %
\def\olafnote#1{\textcolor{magenta}{Olaf:  #1}}
\def\guynote#1{\textcolor{blue}{Guy:  #1}}
\def\luisnote#1{\textcolor{red}{\textbf{Luis:  #1}}}
\def\haitaonote#1{\textcolor{green}{HT:  #1}}

\def\muphys{\bar{\mu}_B}
\def\mubar{\bar{\mu}} %
\def\mubaruvLO{\bar{\mu}_{\tau_F,\mathrm{LO}}} %
\def\mubaruvNLO{\bar{\mu}_{\tau_F,\mathrm{NLO}}} %
\def\mubaruv{\bar{\mu}_{\tau_F}} %
\def\mubarir{\bar{\mu}_{T}} %
\def\mubarirLO{\bar{\mu}_{T,\mathrm{LO}}} %
\def\mubarirNLO{\bar{\mu}_{T,\mathrm{NLO}}} %
\def\muflow{\mu_\mathrm{F}}
\def\muref{\bar{\mu}_\mathrm{ref}} %
\def\murefvalue{2\pi} %
\def\mubaromega{\bar{\mu}_\omega} %
\def\mubaromegaLO{\bar{\mu}_{\omega,\text{LO}}}
\def\mubaromegaNLO{\bar{\mu}_{\omega,\text{NLO}}}
\def\mudr{\bar{\mu}_{\mathrm{DR}}} %
\def\gsquaremsbar{g^2_{\MSBAR}(\mubar)} %
\def\gsquaredflow{g^2_\text{flow}}
\def\alphaflow{\alpha_\text{flow}} %
\def\alphamsbar{\alpha_s} %
\def\flowscale{1/\sqrt{8\tauf}}

\def\GE{G_E}
\def\GB{G_B} %
\def\GBphys{G_B^{\text{phys.}}}
\def\GBflow{G_{B}^{\mathrm{flow}, \muflow}}
\def\GBMSBAR{G_{B}^{\MSBAR,\bar{\mu}}}
\def\GBMSBARmuB{G_{B}^{\MSBAR,\bar{\mu}_B}}
\def\GBMSBARuv{G_{B}^{\MSBAR,\bar{\mu}_{\tau_\mathrm{F}}}} %
\def\GBMSBARir{G_{B}^{\MSBAR,\bar{\mu}_T}} %

\def\Zmatch{Z_{\mathrm{match}}} %

\begin{abstract}
    We analyze the color-magnetic (or ``$B$") field two-point function that encodes the finite-mass correction to the heavy-quark momentum-diffusion coefficient.
    The simulations are done on fine isotropic lattices in the quenched approximation at $1.5\,T_c$, using a range of gradient flow times for noise suppression and operator renormalization.
    The continuum extrapolation is performed at fixed flow time followed by a second extrapolation to zero flow time.
    Perturbative calculations to next-to-leading order of this correlation function, matching gradient-flowed correlators to $\overline{\text{MS}}$, are used to resolve nontrivial renormalization issues.
    We perform a spectral reconstruction based on perturbative model fits to estimate the coefficient $\kappa_B$ of the finite-mass correction to the heavy-quark momentum-diffusion coefficient.
    The approach we present here yields high-precision data for the correlator with all renormalization issues incorporated at next-to-leading order, and is also applicable for actions with dynamical fermions.
\end{abstract}

\maketitle

\section{Introduction}

Heavy-ion collisions create a new state of matter, the quark-gluon plasma
\cite{Collins:1974ky,STAR:2005gfr,BRAHMS:2004adc}.
Studying this state of matter is complicated by the fact that most participants---quarks and gluons---undergo multiple interactions throughout the collision, ending in hadronization.
Therefore, their final yields and momentum distributions encode the early dynamics in a complicated and indirect way.
Hard probes are objects with enough energy and/or momentum to provide more direct information about the early stages of the collision.
Principle among these are heavy quarks, whose evolution in the quark-gluon plasma has been extensively studied both experimentally
\cite{Rapp:2018qla,Dong:2019unq,He:2022ywp}
and theoretically
\cite{Moore:2004tg,Caron-Huot:2007rwy,He:2022ywp}.

Most heavy quarks are produced at relatively low transverse momentum, meaning $\gamma v \lesssim 1$.
In this regime, they interact with the medium through a series of scatterings which can be collectively described as momentum diffusion and drag
\cite{Moore:2004tg,vanHees:2004gq,Mustafa:2004dr}.
Close to rest, the momentum-diffusion coefficient $\kappa$ is determined by a force-force correlation function which can be rewritten as a color-$E$-field color-$E$-field correlation function in the medium, with the adjoint color-electric field group-theory factors connected by a fundamental Wilson line
\cite{CasalderreySolana:2006rq}.
The associated spectral function is the analytical continuation of the thermal Euclidean correlation function of two $E$ fields along a fundamental Polyakov loop, normalized by the trace of the Polyakov loop
\cite{Caron-Huot:2009ncn}.
This opens the possibility of computing this quantity on the lattice, something several groups have pursued
\cite{Francis:2015daa, Altenkort:2020fgs, Brambilla:2020siz,  Banerjee:2022gen, Altenkort:2023oms}.

Because the charm quark is not extremely heavy, and because some charm and bottom quarks are created with somewhat larger momenta, it may also be important to consider finite-velocity corrections to this momentum-diffusion picture.
Bouttefeux and Laine showed
\cite{Bouttefeux:2020ycy}
that the next order in velocity is determined by a color-magnetic-magnetic correlator on a Wilson line, which is the continuation of a magnetic-magnetic correlator along a fundamental Wilson line in the thermal Euclidean path integral.
This quantity can also be investigated on the lattice, and several groups have already tried to do so
\cite{Banerjee:2022uge, Brambilla:2022xbd,Altenkort:2023eav}.
However, treating this problem is more complicated, as Laine already showed
\cite{Laine:2021uzs}:
the $BB$ correlator on a Wilson line renormalizes in a nontrivial way, which must be taken into account both in fitting lattice data and in interpreting the result in terms of heavy-quark diffusion.

The goal of this paper is to investigate the correlation function of two magnetic fields on a Wilson line with the use of lattice QCD and gradient flow.
We hope to improve on previous studies in three ways.
First, we perform a high-statistics study using a wide range of very fine lattices in order to control continuum and small-flow extrapolations.
Second, we treat correlations within the data analysis process more carefully than previous studies (including our previous study of electric field correlations).
For instance, results at a fixed separation and nearly equal flow depths are highly correlated, whereas many analyses have treated different flow depths as if they are independent; similarly, the small-flow extrapolation for nearly equal separations is also highly correlated.
Third, we treat the issues of operator renormalization more carefully when carrying out the finite flow time extrapolation.
Here we take advantage of a recent next-to-leading order (NLO) result for the effect of gradient flow on magnetic field correlators along a Wilson line to provide a fully NLO matching between gradient-flowed lattice results, perturbative calculations in the modified minimal subtraction scheme ($\MSBAR$ scheme), and the physical diffusion coefficient.
The use of these relations greatly improves the extrapolation over gradient flow depth, eliminating a logarithmic dependence which would otherwise arise.

An outline of the paper is as follows.
In \cref{sec:renorm} we present the theoretical framework of this study, including the definition of lattice object, analytical continuation, gradient flow and a matching procedure for converting lattice results at finite gradient flow depth to the physical values.
Next, \cref{sec:latt} describes the details of our lattice setup and the evaluation of magnetic-field correlation functions.
\cref{sec:coupling} provides the details of the matching procedure.
\cref{sec:extrap} describes the continuum and zero-flow limits.
This section represents a main innovation in our treatment, since the renormalization of the operators as a function of gradient flow depth is fully incorporated in our extrapolations, and since data correlations across flow times and between separations is taken into account.
Then \cref{sec:continue} presents the analytical continuation of our results to a Minkowski spectral function.
Finally, \cref{sec:results} presents our results and conclusions.

\section{Renormalization of magnetic-field correlators under gradient flow}
\label{sec:renorm}

The basic object we shall consider is the correlation function of two magnetic field operators along a Wilson line.
In Euclidean space this will be a fundamental-representation Polyakov loop, and the value should be normalized by the mean trace of the Polyakov loop:
\begin{align}
    \label{Geucl}
    G_B(\tau)  & \equiv L^{-1} \left\langle \Tr G_{12}(0) U(0,\tau) G_{12}(\tau) U(\tau,\beta)  \right\rangle ,
    \\
    \label{Ldef}
    L          & \equiv \left\langle \Tr U(0,\beta) \right\rangle ,
    \\
    \label{Wilsonlinedef}
    U(t_1,t_2) & \equiv \mathrm{Pexp} -i\int_{t_1}^{t_2} G_0^A(t) T^A dt \,.
\end{align}
Here $\beta = 1/T$ is the inverse temperature, $U(t_1,t_2)$ is a fundamental-representation Wilson line along the time direction, $G_{12}$ is the color magnetic field strength under the geometrical normalization convention, and $L$ is the trace of the Polyakov line.
All operators occur at a common spatial point, only time is varying.
The analytical continuation of this object to real time is the Wightman-ordered correlation function of two magnetic field operators with a Wilson line running up the real time axis, between the operators, down the real-time axis, and across the Euclidean branch:
\begin{align}
    \label{Gmink}
    G_B(t) & = L^{-1} \left\langle \Tr G_{12}(0) U(0,t) G_{12}(t)
    U(t,0) U(0,i\beta) \right\rangle .
\end{align}
The associated spectral function $\rho(\omega)$ is related to $G_B(\tau)$ in the standard way,
\begin{equation}
    \label{continuation}
    G_B(\tau) = \int_0^\infty \frac{d\omega}{\pi} \rho_B(\omega)
    \frac{\cosh\left(\omega\left(\tau-\beta/2\right)\right)}
    {\sinh(\omega\beta/2)} \,.
\end{equation}
In a completely analogous way one can also define $\GE$, the correlation function for electric fields,
\begin{equation}
    \label{GEdef}
    \GE(\tau) = -L^{-1} \langle \Tr G_{01}(0) U(0,\tau) G_{01}(\tau) U(\tau,\beta) \rangle,
\end{equation}
where the $-$ sign accounts for two factors of $i$ in the continuation of Minkowski to Euclidean electrical fields.%
\footnote{$\GE(\tau)$ defined in this way is positive because the Euclidean electric field is time-reflection odd.}
The two spectral functions $\rho_E$ and $\rho_B$ then determine the momentum-diffusion coefficient $\kappa$ via
\begin{equation}
    \label{kappa}
    \kappa = \lim_{\omega \to 0} \frac{2T}{\omega}
    \left( \rho_E + \frac{2}{3} \langle \mathbf{v}^2 \rangle \rho_B \right),
\end{equation}
where $\langle \mathbf{v}^2 \rangle$ is the mean-squared velocity of the heavy quark. The velocity, as determined by thermodynamics, depends on the ratio $T/M$, which is the way the dependence on the thermal heavy quark mass $M$ enters.
The operators in \cref{Geucl}, \cref{Gmink} require regularization and their regularized values will generically depend on the scale or procedure used.
For the case of electric field correlators on a Wilson line this regularization turns out to be harmless;
the operator determining $\GE$ turns out to be finite and regulator-independent within $\MSBAR$ and gradient flow regulators, and its value in these two schemes is equal.
But the issue is more complicated for $G_B(\tau)$, as we review below.

The lattice itself provides a regularization which could be used.
But the regularization calculation has not been carried out to our knowledge, and in general lattice-regularized operators tend to possess large renormalizations and poorly convergent perturbative expansions, so this approach is not attractive.
In any case, achieving a high signal to noise from a lattice simulation will generally require the use of gradient flow.
Gradient flow automatically provides renormalized operators with finite correlation functions, and it eliminates any detailed dependence on the lattice spacing or lattice implementation.
But it is important to correctly understand what this regularization actually does in terms of the flow time dependence of $G_B$.

Briefly, gradient flow is a procedure for transforming gauge-field backgrounds $A_\mu(x)$ within the path integral, removing their short-distance fluctuations.
One defines a flowed variable $B_\mu(x,\tauf)$ through a boundary condition and an evolution equation,
\begin{align}
    \label{flowdef}
    B_\mu(x,\tauf=0)                & = A_\mu(x) \,, \nonumber
    \\
    \partial_{\tauf} B_\mu(x,\tauf) & = D_\nu^B G_{\nu\mu}^B ,
\end{align}
where the superscript $B$ indicates that the covariant derivative and field strength are those constructed using the flowed field $B_\mu(x,\tauf)$, not the original field $A_\mu(x)$.
For more discussion see \ccites{Narayanan:2006rf,Luscher:2009eq,Luscher:2010iy,Luscher:2011bx}.
After applying this procedure to a flow depth $\tauf$, the gauge fields are free of short-distance fluctuations beyond approximately the momentum scale%
\footnote{Note that $\tauf$ has units of length squared, not length.}
\begin{equation}
    \label{muflowdef}
    \muflow \equiv \flowscale
\end{equation}
and all correlation functions are rendered finite.
We will write the magnetic field correlation function evaluated on the flowed fields as $\GBflow$,
\begin{equation}
    \label{GBflowdef}
    \GBflow(\tau) \equiv G_B(\tau) \; \mbox{with} \;
    A_\mu(x) \to B_\mu(x,\tauf) \,.
\end{equation}
It is the quantity which one directly measures on the lattice.
The quantity $\GE$ has a finite $\tauf \to 0$ limit which is approached with $\OO(\tauf/\tau^2)$ corrections.
However, as we will see, $\GBflow$ also contains logarithmic-in-$\tauf$ corrections which must be correctly handled.

Bouttefeux and Laine \cite{Bouttefeux:2020ycy}, followed by Laine \cite{Laine:2021uzs} considered $G_B$ in the $\MSBAR$ renormalization scheme and its relation to the physical diffusion coefficient for heavy quarks.
They found the relation between the physical diffusion coefficient and the $\MSBAR$-regulated and renormalized%
\footnote{Meaning that $1/\bar\epsilon$ factors have been subtracted off.}
correlation function $\GBMSBAR$ to be
\begin{equation}
    \label{phys-to-MS}
    \GBphys = \GBMSBAR \left( 1 + g^2 \gamma_0 \left[
        \ln \frac{\mubar^2}{(4\pi T)^2} - 2 + 2 \gammaE
        \right] \right) .
\end{equation}
Here $g^2 \gamma_0 = g^2 C_A/8\pi^2 = 3g^2/8\pi^2$ is the leading-order anomalous dimension for a magnetic field operator inserted along a temporal Wilson line (with $C_A=3$ the adjoint Casimir) with the gauge coupling $g^2=4\pi \alphas$.
We assume that $g^2 = \gsquaremsbar$.
Since $\GB$ involves two $B$-field insertions, the correction has an additional overall factor of $2$ relative to that in \ccite{Laine:2021uzs}.
This expression implies that $\GBMSBAR$ depends explicitly on the $\MSBAR$ renormalization scale $\mubar$ through a Callan-Symanzik equation:
\begin{align}
    \label{CallanSymanzik}
    \frac{\mubar^2 \mathrm{d}}{\mathrm{d}\mubar^2} \GBphys  & = 0 \nonumber                \\
    \Rightarrow \;\;
    \frac{\mubar^2 \mathrm{d}}{\mathrm{d}\mubar^2} \GBMSBAR & = - g^2 \gamma_0 \GBMSBAR\,.
\end{align}
It remains to relate $\GBMSBAR$ with what we actually measure on the lattice, which is $\GBflow$.
Since the latter is independent of the $\mubar$ scale, its relation with $\GBMSBAR$ must also involve logarithms of $\mubar$, balanced by the gradient flow scale $\muflow$.
Very recently, de la Cruz \textit{et al.} have found the relation \cite{delacruzinprep}
between these quantities, valid to leading order in small $\tauf/\tau^2$,
\begin{equation}
    \label{MS-to-flow}
    \GBflow = \GBMSBAR \left( 1 + \gamma_0 g^2 \left[
        \ln \frac{\mubar^2 }{4\muflow^2} + \gammaE
        \right] \right).
\end{equation}
This result was also recently derived by Brambilla and Wang
\cite{brambilla2023offlightcone}.
The $\mubar$ dependence in \cref{MS-to-flow} and in \cref{phys-to-MS} cancel, rendering a combined result which is independent of $\mubar$ at the NLO level.

Besides these renormalization effects, we expect that $\GBflow$ will contain additional polynomially suppressed flow effects, that is, effects of order $\tauf/\tau^2$.
To extrapolate these away, we will be interested in small values of $\tauf$, satisfying $\tauf T^2 \ll 1$ and therefore $\muflow \gg T$.
If we use a single $\MSBAR$ scale $\mubar$ in applying both \cref{phys-to-MS,MS-to-flow}, this will lead to large logarithms in at least one equation.
We should then expect $\alphas^2 \ln^2(\mathrm{large})$ effects at the next order.
Such uncontrolled higher-order logarithms can be avoided, as usual, by evaluating \cref{phys-to-MS,MS-to-flow} at two different $\MSBAR$ scales, each chosen appropriately to avoid large logarithms in the matching, and evolving $\GBMSBAR$ between these scales using the Callan-Symanzik equation \cref{CallanSymanzik}.
This also requires an accurate determination of the scale-dependent gauge coupling.
Combining these ideas, the following approach appears appropriate to obtain the renormalized color-magnetic correlator $\GBphys$:
\begin{enumerate}
    \label{recipes}
    \item \label{item1} Determine the $\MSBAR$ gauge coupling $\gsquaremsbar=4 \pi \alpha_s$ at one scale, $\muref$, accurately, using the approach of
          \ccites{Luscher:2010iy,Harlander:2016vzb} (more details in \cref{sec:coupling}).
          From this, use the standard $\MSBAR$ beta function to determine the gauge coupling over the desired range of scales.
    \item \label{item2} For each relevant\footnote{See \cref{sec:extrap} for details. Relevant flow times are some set of flow times inside the flow time extrapolation window. In this window, data at different flow times are highly correlated. To not underestimate statistical errors, we perform the flow time extrapolation of $\GBflow$ using as few points as possible (i.e., three maximally spaced data points for a linear fit).} flow time, measure $\GBflow(\tauf)$ on the lattice and take its continuum limit at fixed flow time.
    \item \label{item3} Match the continuum-extrapolated $\GBflow$ to  $\GBMSBAR$ with the help of \cref{MS-to-flow}.
          This requires choosing a matching scale $\mubaruv$ for $\mubar$.
          This could be, for example,
          \begin{align}
              \mubaruvLO \equiv \muflow,
          \end{align} the typical scale for gradient flow effects, or the scale where $\GBflow=\GBMSBAR$, which is
          \begin{align}
              \mubaruvNLO \equiv \muflow \sqrt{4e^{-\gammaE}} \simeq 1.50\,\muflow.
          \end{align}In any case we end up with $\GBMSBARuv$.
    \item \label{item4}
          To avoid large logarithms in \cref{phys-to-MS}, run $\GBMSBARuv$ to the ``physical scale'' $\mubarir$ using \cref{CallanSymanzik}, incorporating the running of the coupling using a two-loop or higher-loop beta function.
          Two sensible choices for $\mubarir$ would be
          \begin{align}
              \mubarirLO \equiv 2\pi T,
          \end{align} the typical scale for thermal physics, and the scale where $\GBphys=\GBMSBAR$ at NLO, which is
          \begin{align}
              \mubarirNLO \equiv 4\pi e^{1-\gammaE} T \simeq 19.18 T.
          \end{align} In any case we end up with $\GBMSBARir$.
    \item \label{item5}
          Finally, with the help of \cref{phys-to-MS}, obtain $\GBphys$ from $\GBMSBARir$.
\end{enumerate}
In the event that either \cref{phys-to-MS} or \cref{MS-to-flow} involves a relatively large correction, we can also attempt to incorporate higher-loop multiple-log contributions by the substitution $(1+x) \to \exp(x)$, where $x=\gamma_0 g^2 \ldots$ is the NLO correction.
A summary of steps \ref{item3}--\ref{item5} is then
\begin{align}
    \label{eq:Ztotal}
    \GBphys(\tauf) = {} & \Zmatch(\mubarir,\mubaruv,\muflow) \; \GBflow(\tauf) \,,
    \nonumber                                                                                                      \\
    \ln \Zmatch = {}    & \int^{\mubaruv^2}_{\mubarir^2}\gamma_0 \gsqms(\mubar)
    \frac{d\mubar^2}{\mubar^2}                                                                                     \\
                        & + \gamma_0 \gsqms(\mubarir) \left[
        \ln \frac{\mubarir^2}{(4\pi T)^2} - 2 + 2 \gammaE \right]
    \nonumber                                                                                                      \\ \nonumber
                        & - \gamma_0 \gsqms(\mubaruv) \left[ \ln\frac{\mubaruv^2}{4\,\muflow^2} + \gammaE \right],
\end{align}
where we have introduced the total matching factor $\Zmatch$.
The resulting quantity, $\GBphys(\tauf)$, must then be extrapolated to zero flow time, assuming that the flow time dependency is of form $\tauf/\tau^2$ respectively.

In the following we will employ this procedure in our data analysis to correctly account for the renormalization of $\GB$, such that the spectral reconstruction can be performed using $\GBphys$.
The approach involves a somewhat arbitrary choice for the scale $\mubarir$ and for the scale $\mubaruv$.
The dependence on these choices formally cancels at the NLO level, but it will introduce residual next-to-next-to-leading order (NNLO) effects.
By considering two somewhat different choices for each scale, we will determine the residual renormalization-point dependence of our results, which we will include in our systematic error budget.

\section{Lattice details}
\label{sec:latt}
\begin{figure*}
    \null \hfill
    \includegraphics[scale=1]{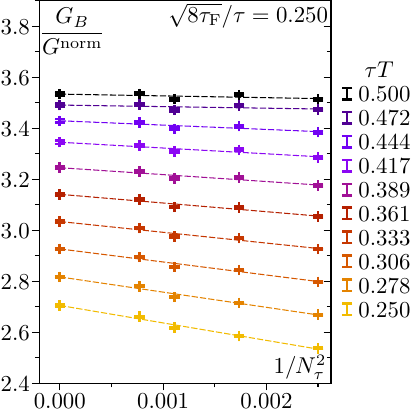}
    \hspace{2.1cm}
    \includegraphics[scale=1]{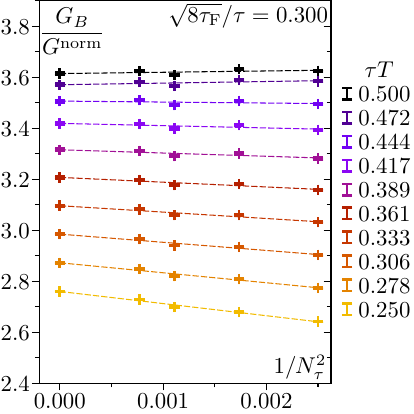}
    \hfill \null
    \caption{Bare color-magnetic correlator $G_B$, tree-level-improved and normalized to its free counterpart $G^\text{norm}$, as a function of squared lattice spacing $a^2$ (or equivalently $1/N_\tau^2$ at fixed temperature $T=1/(a N_\tau)$) at the smallest (left) and largest (right) flow time in units of $\sqrt{8\tauf}/\tau$ according to \cref{eq:flow-extr-window}. The dashed lines and data points at $1/N_\tau^2=0$ represent the linear-in-$a^2$ continuum extrapolation. Statistical errors are smaller than the error bar linewidth.}
    \label{fig:corr-cont-extrapo}
\end{figure*}

The color-magnetic field correlators are measured on four isotropic lattices at
temperature $T \simeq 1.5 T_c$, where
$T_c$ is the confinement/deconfinement phase transition temperature determined via the Sommer parameter $r_0$ \cite{Sommer:1993ce} in \ccite{Francis:2015lha}.\footnote{
    The gauge configuration ensemble, detailed in \cref{tab:lattice_setup}, was generated years ago with parameters not precisely tuned to $1.5 T_c$. These discrepancies, possibly impacting the continuum extrapolation slightly (cf.\ \cref{fig:corr-cont-extrapo}), are ignored for the purpose of this study.}
The gauge configurations are generated with the standard Wilson gauge action using the heat bath and overrelexation algorithms.
To ensure our calculation is performed on fully thermalized configurations, the first 4000 sweeps, each consisting of one heat bath update and four overrelexation updates, have been discarded.
Autocorrelation between measurements was eliminated by sampling only every 500th sweep.
Periodic boundary conditions are used for all directions.
The lattice spacing $a$ is determined via the $r_0$ scale \cite{Francis:2015lha, Burnier:2017bod}.
The finite temperature lattice setup is summarized in \cref{tab:lattice_setup}. For the calculation of the coupling constant, we use additional zero-temperature lattice simulations at the three finest lattice spacings considered here, see \cref{tab_zeroT}.

\begin{table}[t]
    \centering
    \begin{tabular}{ccrcccc}
        \hline \hline
        $a$ (fm) & $a^{-1}$ (GeV) & $N_{\sigma}$ & $N_{\tau}$ & $\beta$ & $T/T_{c}$ & No. configuration\tabularnewline
        \hline
        0.0215   & 9.187          & 80           & 20         & 7.0350  & 1.4734    & 10000 \tabularnewline
        0.0178   & 11.11          & 96           & 24         & 7.1920  & 1.4848    & 10000 \tabularnewline
        0.0140   & 14.14          & 120          & 30         & 7.3940  & 1.5118    & 10000 \tabularnewline
        0.0117   & 16.88          & 144          & 36         & 7.5440  & 1.5042    & 10000 \tabularnewline
        \hline \hline
    \end{tabular}
    \caption{ The lattice setup used for the calculation of the color-magnetic field correlators. Here, $\beta$ is the inverse gauge coupling, not to be confused with the inverse temperature defined below (\cref{Geucl}).}
    \label{tab:lattice_setup}
\end{table}

\begin{table}[t]
    \centering
    \begin{tabular}{ccrccccc}
        \hline \hline
        $a$ (fm) & $a^{-1}$ (GeV) & $N_{\sigma}$ & $N_{\tau}$ & $\beta$ & $T/T_{c}$ & No. configuration\tabularnewline
        \hline
        0.0178   & 11.11          & 96           & 96         & 7.1920  & 0.3712    & 1000 \tabularnewline
        0.0140   & 14.14          & 96           & 120        & 7.3940  & 0.3780    & 1000 \tabularnewline
        0.0117   & 16.88          & 96           & 144        & 7.5440  & 0.3761    & 1000 \tabularnewline
        \hline \hline
    \end{tabular}
    \caption{ The lattices at $T<T_c$ (``zero temperature'') for determining the gauge coupling.
    }
    \label{tab_zeroT}
\end{table}

\begin{figure*}[t]
    \null \hfill
    \includegraphics[scale=1]{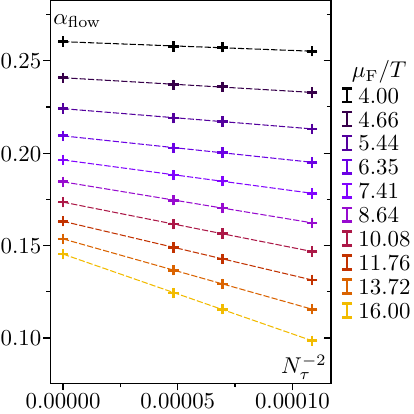}\hspace{2.1cm}
    \includegraphics[scale=1]{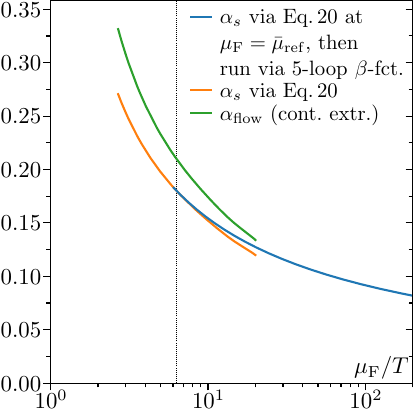}\hfill \null
    \caption{Left: gradient flow coupling $\alphaflow$ as a function of squared lattice spacing [or equivalently $N_\tau^{-2}$ at fixed temperature $T=1/(a N_\tau)$] at selected flow scales $\muflow$ in temperature units. The dashed lines correspond to linear fits at fixed flow scale, yielding the continuum-extrapolated data points at $N_\tau^{-2}=0$. Statistical errors are smaller than the data linewidths. Right: continuum-extrapolated flow coupling $\alphaflow$ and matched $\MSBAR$ coupling $\alphamsbar$ (via \cref{harlander}) as a function of gradient flow scale $\muflow$ in temperature units. Note that the couplings and scales are defined such that $\muflow=\mubar$. The dotted vertical line is located at $\muflow = \muref$ (\cref{eq:muref}) and depicts the point from where we move to other scales via the perturbative beta function. Statistical errors are smaller than the linewidth.}
    \label{fig:g2-cont-extrap}
\end{figure*}

Before measuring the color-magnetic field correlator on the lattice, we first evolve the gauge fields to the desired flow time range using the Symanzik-improved gradient flow (Zeuthen flow) \cite{Ramos:2015baa}.

The generation of gauge configurations and the correlator measurements, as well as the gradient flow evolution is performed using the \texttt{SIMULATeQCD} suite \cite{Mazur:2023lvn,Altenkort:2021cvg,Mazur:2021zgi}. The correlator data analysis and spectral reconstruction is carried out using our software toolkit \texttt{correlators\_flow}~\cite{BBdatapublication}.

\section{Determination of gauge coupling and matching factor}\label{sec:coupling}

In order to make use of \cref{eq:Ztotal} and the matching procedure explained in \cref{sec:renorm}, the gauge coupling has to be known in the range
from $\mubaruv$ to $\mubarir$.
We can determine the $\MSBAR$ gauge coupling $g^2_{\MSBAR}=4\pi \alphamsbar$ self-consistently from the lattice data over a range of scales from about $\mu=1/a$ to about $\mu=\pi T$ by evaluating the squared field strength $E= \Tr G_{\mu\nu} G_{\mu\nu}/2$ under gradient flow, which yields
\begin{equation}
    \label{ggfdef}
    \alphaflow (\muflow) \equiv \frac{4\pi}{3}  \tauf^2 \langle E \rangle_{\tauf} \,.
\end{equation}
Here, $\displaystyle \langle E \rangle_{\tauf}$ is the expectation value of the squared field strength $E$ at flow time $\tauf$.
Then we use the relation between this quantity and the $\MSbar$ coupling, calculated at NLO by L\"uscher \cite{Luscher:2010iy} and at NNLO by Harlander and Neumann \cite{Harlander:2016vzb}:
\begin{align}
    \label{harlander}
    \alphaflow & = \alphamsbar
    \left( 1 + k_1 \alphamsbar + k_2 \alphamsbar^2 \right)
    \quad \mbox{for} \; \mubar^2  = \muflow^2 \,,
    \\
    k_1        & = 1.098 + 0.008 N_f \,, \nonumber                \\
    k_2        & = -0.982 - 0.070 N_f + 0.002 N_f^2 \,. \nonumber
\end{align}
The $\MSBAR$ coupling $\alphamsbar$ is then obtained by solving the cubic equation. Note that $N_f$ is the number of light quark flavors which is zero in the case of this study.
This approach becomes unreliable for $\tauf < a^2$ (where $a$ means lattice spacing) due to missing $\OO(a^2/\tauf)$ effects.
It also has problems when $\tauf$ becomes too large, as the matching then occurs in a stronger-coupled regime where the perturbative match of \cref{harlander} becomes less precise.
This makes it difficult to establish $\alphamsbar$ directly over the full range from $\mubarir$ to $\mubaruv$.
Therefore the best procedure appears to be to use this direct determination at a single value $\muflow=1/\sqrt{8\tauf}=\muref$ where both lattice spacing and perturbative effects are under control, and to use the perturbative running of the gauge coupling to obtain it in the desired range.
This is the approach we will follow in this work.

\begin{figure*}[t]
    \null \hfill
    \includegraphics[scale=1]{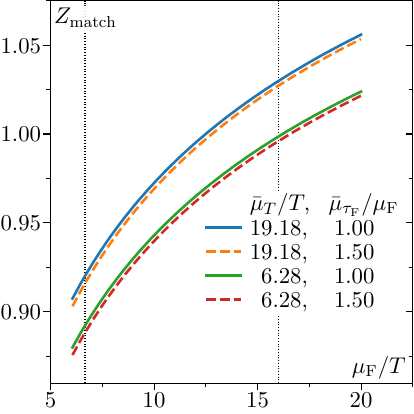}
    \hspace{2.1cm}
    \includegraphics[scale=1]{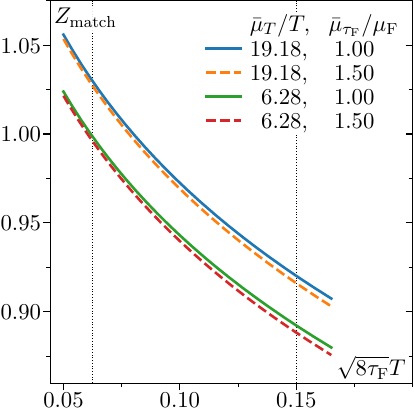}
    \hfill \null
    \caption{The renormalization factor of the color-magnetic correlator, $\Zmatch$ (\cref{eq:Ztotal}), as a function of flow scale $\muflow$ in temperature units (left) or as a function of flow radius $\sqrt{8\tau_F}$ in inverse temperature units (right) for the four combinations of $\mubarir$ and $\mubaruv$ defined in \cref{sec:renorm}. Statistical errors are smaller than the linewidth. The difference between the panels is the inverted x-axis since $\muflow= 1/(\sqrt{8\tauf})$.
    }
    \label{fig:Zmatch}
\end{figure*}

Since the matching procedure is conducted based on the continuum theory, it is necessary to extrapolate the flowed coupling constant $\alpha_\mathrm{flow}(\muflow)$ measured on the lattice (see \cref{tab_zeroT}) to the continuum first.
Given that the discretization error of the lattice action is of order $a^2$, we adopt an \textit{Ansatz} linear in $a^2$.
This extrapolation is shown in the left panel of \cref{fig:g2-cont-extrap}.
It can be seen that the \textit{Ansatz} describes the lattice data well for all relevant flow times. However, the figure shows that the continuum correction is only small for $\muflow\lesssim 8T$, correspondingly restricting the choice for $\muref$.
In the right panel of \cref{fig:g2-cont-extrap}, the continuum-extrapolated (cont.~extr.) $\alpha_\mathrm{flow}(\muflow)$ is shown as a solid green curve. Converting it to $\alphas$ via \cref{harlander} yields the solid orange curve.
We select
\begin{align}
    \muref \equiv \murefvalue T \label{eq:muref}
\end{align}
and have checked that reducing this value by half does not significantly alter the results.
From this point we can move to other scales via the perturbative beta function.
In practice, the evolution is calculated using the function \texttt{crd.AlphasExact($\alpha_s(\muref)$, $\muref$, $\mubaruv$, $N_f=0$, $N_{\mathrm{loop}}=5$)} of the RunDec package \cite{Herren:2017osy, Chetyrkin:2000yt}.
The resulting coupling is then used to compute $\Zmatch$ via \cref{eq:Ztotal}, which we show in \cref{fig:Zmatch} for different choices of $\mubarir$ and $\mubaruv$.
The vertical dotted lines enclose all flow times we will use when extrapolating to zero flow (the flow time extrapolation window), meaning that the renormalization procedure changes the bare correlator by at most between $-15\%$ and $+5\%$. We can also see that the difference between choices for $\mubaruv$ is almost negligible, while for $\mubarir$ the difference is simply a constant multiplicative factor.
The four curves for $\Zmatch$ shown in this figure, which formally differ from each other at the NNLO level, will be used in the next section to obtain four physical correlators via \cref{eq:Ztotal}.
These will be treated on an equal footing in the spectral analysis and therefore contribute to the overall systematic uncertainty.
However, as we will see later, the impact of the different choices on the determination of $\kappa_B$ is almost negligible.
For this reason, some figures are simplified to show only the choice $(\mubaruv = \mubaruvNLO;\: \mubarir=\mubarirNLO)$.

\section{Correlator computation}
\label{sec:extrap}

We calculate the color-magnetic correlator $G_B$ on the lattice using four different lattice spacings $a$ (cf.\ \cref{tab:lattice_setup}) and perform bootstrap resampling to carry out the complete analysis procedure (i.e., continuum extrapolation, renormalization and flow time extrapolation of $G_B$, and in the next section spectral reconstruction) on 1000 bootstrap samples.\footnote{For each lattice spacing, 1000 samples for the correlator are created by randomly drawing 10,000 measurements at that lattice spacing with replacement, from which then, for each flow time, the corresponding sample correlator is calculated. The continuum extrapolation yields 1000 continuum correlator samples by drawing one sample correlator from each lattice spacing (without replacement). We represent bootstrap sample distributions using their median $\pm 34\mathrm{th}$ percentile, which is what is shown in all subsequent figures that feature data points with error bars.}

Note that to suppress discretization errors, we tree-level improve \cite{Meyer:2009vj} the bare correlator by multiplying it with the ratio of the free correlator obtained in the continuum at leading order $G_B^{\mathrm{norm}}(\tau)$ \cite{Caron-Huot:2009ncn} (which is same as in the color-electric case) and the one calculated in lattice perturbation theory at leading order at finite flow time \cite{Altenkort:2023oms}
\begin{equation}
    G_B^{\rm{latt}}(\tau,\tauf) \rightarrow  G_B^{\rm{latt}}(\tau,\tauf) \times
    \frac{G_B^{\mathrm{norm}}(\tau)}{G_B^\stackss{norm}{latt}(\tau, \tauf)}.
\end{equation}
This is equivalent to including tree-level lattice-spacing errors in both the correlator and gradient flow procedures.
To make the subtle features of the data visible, we further normalize it by $G_B^{\mathrm{norm}}(\tau)$. The quantity of interest is therefore the tree-level-improved ratio
\begin{align}
    \label{Gimp}
    \frac{G_B^{\rm{latt}}(\tau,\tauf)}{G_B^\stackss{norm}{latt}(\tau, \tauf)}.
\end{align}
For simplicity we drop the trivial superscript ``latt" in the following.

\subsection{Continuum extrapolation of the bare correlator}

In this section we consider the continuum extrapolation of the bare correlators (\cref{item2} of \cref{sec:renorm}). The continuum extrapolation of $G_B(\tau,\tauf)/G_B^\mathrm{norm}(\tau, \tauf)$ requires an interpolation of the Euclidean temporal separations $\tau T$ to the same values across all lattices (we use those present on the finest lattice). This is done using cubic splines with natural boundary condition at the smallest $\tau T$ and zero slope at $\tau T=0.5$ due to the periodic boundary conditions used in the lattice simulation. The continuum extrapolation is carried out using an \textit{Ansatz} linear in $a^2$ as we use the standard Wilson action.

\begin{figure*}
    \centerline{
        \null \hfill
        \includegraphics[scale=1]{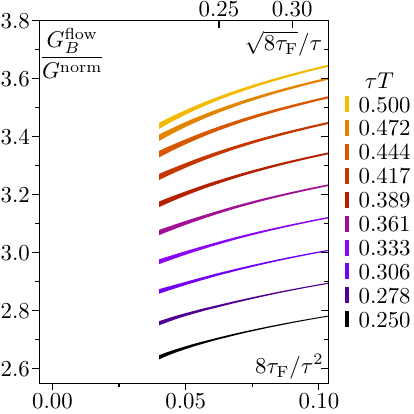}\hspace{2.1cm}
        \includegraphics[scale=1]{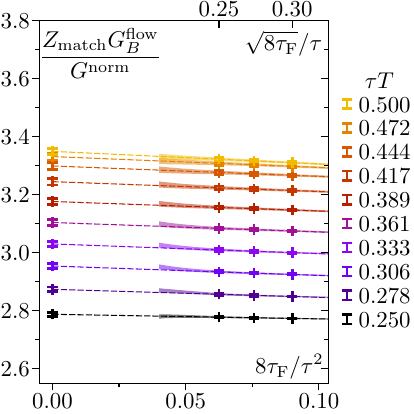}
        \hfill\null
    }
    \caption{Left: bare continuum-extrapolated color-magnetic correlator $G_B$, tree-level-improved and normalized to its free counterpart $G^\text{norm}$, as a function of normalized flow time $\tauf/\tau$ for Euclidean temporal separations $\tau$ in units of inverse temperature. The bands depict the statistical $\pm 1 \sigma$ error around the median value. Right: the same as on the left, but renormalized by multiplying with $\Zmatch$ (\cref{eq:Ztotal}) for the case $\mubarir=\mubarirNLO$, $\mubaruv=\mubaruvNLO$ (other choices differ only marginally or by a simple rescaling). The dashed lines depict the linear-in-$\tauf/\tau^2$ flow time extrapolation, which, due to large autocorrelation between subsequent flow times, is based solely on the three explicit data points indicated between $0.25 \leq \sqrt{8\tauf}/\tau \leq 0.30$ (\cref{eq:flow-extr-window}).}
    \label{fig:flow-extrapo}
\end{figure*}

In the two panels of \cref{fig:corr-cont-extrapo} we show the extrapolation at the minimum and maximum flow time [in units of $\sqrt{8\tauf}/\tau$, cf.\ \cref{eq:flow-extr-window}] that is later used in the flow-time-to-zero extrapolation. We can see that the linear-in-$a^2$ fit \textit{Ansatz} gives a good description for the lattice data at all $\tau T$ considered (smaller $\tau T$ are not reliably accessible by the gradient flow method, as explained further below). The continuum-extrapolated results are shown in the left panel of \cref{fig:flow-extrapo}. The renormalized correlators obtained via \cref{eq:Ztotal} are shown in the right panel of \cref{fig:flow-extrapo} as colorful bands. It can be seen that the matching factor does ameliorate the logarithmic flow time divergence, rendering the remaining flow time dependence of the renormalized color-magnetic correlator linear, which is in close resemblance to the behavior of the color-electric correlator (cf.\ \ccites{Altenkort:2020fgs, Altenkort:2023oms}).

\subsection{Flow-time-to-zero extrapolation of renormalized correlator}

\begin{figure}[b]
    \centering
    \includegraphics[scale=1]{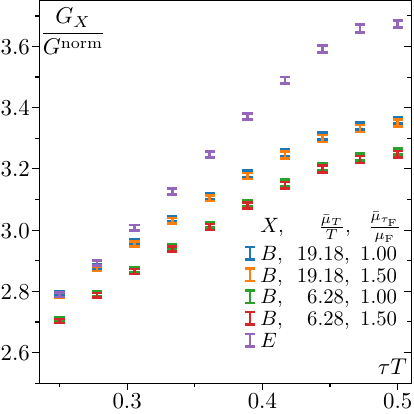}
    \caption{Comparison of renormalized, continuum, zero flow time color-electric ($X=E$, from \ccite{Altenkort:2020fgs}) and color-magnetic ($X=B$, this work) correlators as a function of Euclidean temporal separation $\tau$ in inverse temperature units. For the color-magnetic case four variants are shown that arise from the renormalization due to \textit{a priori} unknown scales at NLO as detailed in \cref{sec:renorm}.
    }
    \label{fig:EEvsBB}
\end{figure}

Now we perform the $\tauf\rightarrow 0$ extrapolation of the continuum color-magnetic correlator, which has been renormalized via \cref{eq:Ztotal} using the matching factors shown in \cref{fig:Zmatch}.
As detailed in a previous study of the color-electric field correlators \cite{Altenkort:2020fgs, Altenkort:2023oms}, there is a small flow time window that is usable for the flow extrapolation, which reads:
\begin{align}
    0.25 \leq \frac{\sqrt{8\tauf}}{\tau} \leq 0.3.
    \label{eq:flow-extr-window}
\end{align}
The lower limit effectively results from the largest lattice spacing used in the continuum extrapolation, as enough flow needs to be applied to sufficiently suppress lattice discretization artifacts. The upper limit results from a perturbative study \cite{Eller:2018yje} that allowed the flowed (color-electric) correlator to deviate from its zero flow counterpart by, at most, 1\%. This ensures that the flowed correlator does not stray too far from the physical, zero flow time value. Due to \cref{eq:flow-extr-window}, the correlator can only be calculated reliably for Euclidean time separations $\tau T \geq 0.25$.

In the window given by \cref{eq:flow-extr-window} a linear extrapolation is performed, inspired by the lowest-order small flow time expansion and justified by our data.
Such a simple \textit{Ansatz} turns out to work very well as the right panel of \cref{fig:flow-extrapo} indicates, where the fits, indicated by the dashed lines, go through all data points in the flow time window.
We perform this extrapolation by fitting to the three explicitly shown data points at the minimum, midpoint, and maximum of our flow time window.
We do this because there is a large autocorrelation between nearby flow times.
Fitting to many nearby flow times is therefore not justified statistically.
It is also important that we determine the error in the fit based on the different fit results found in bootstrap resamplings of our data, rather than through naive regression.

In \cref{fig:EEvsBB} we show a comparison of the double-extrapolated color-electric field correlator $G_E$ from \ccite{Altenkort:2020fgs} and the color-magnetic field correlators $\GBphys$ obtained in this study at the same temperature using the same lattices.
The figure shows that each correlator is larger at $\tau T = 0.5$ than at small $\tau T$ values.
This is partly a result of infrared, thermal contributions to the spectral function, and partly because of the intrinsic separation dependence of the vacuum correlation function.
For $G_E$ these arise because $G_E \propto g^2$ which is scale dependent and becomes smaller at short distances---hence the falloff in $G_E$ between, say, $\tau T = 0.35$ and $\tau T = 0.25$.
For $G_B$ this distance dependence arises from $g^2$ and also from the scale (and hence $\tau$) dependence of $\Zmatch$, or equivalently, of $c_B^2$ as we will discuss around \cref{phiUV2}.
This softens the $\tau$ dependence of the vacuum $G_B$ correlator in comparison to $G_E$.

\section{spectral reconstruction}
\label{sec:continue}

In this section we invert \cref{continuation} to reconstruct the spectral function from the renormalized color-magnetic field correlators $\GBphys$ obtained above.  We model the spectral function consisting of two main parts: the ultraviolet (UV) part and the infrared (IR) part. The UV part is known at both leading order (LO) \cite{Banerjee:2022uge} and next-to-leading order (NLO) \cite{Banerjee:2022uge}. At LO the $\MSBAR$ spectral function matches the color-electric case and reads
\begin{align}
    \phi^{\mathrm{LO}}_{\rm{UV}}(\omega; \mubaromega) = \frac{g^2(\bar{\mu}_\omega) C_F \omega^3}{6\pi},
    \label{phiUV1}
\end{align}
with $C_F=(N_c^2-1)/(2N_c)$, $N_c=3$, and $g^2$ the $\MSBAR$ coupling, which should be evaluated at some scale which may be $\omega$ dependent.
In a previous paper \cite{Altenkort:2023eav} we use this LO spectral function for the UV part, picking a renormalization point which transitions from a fixed value in the IR to $\mubar = 2\omega$ in the UV.
However, this approach is not really self-consistent, since it takes into account one source of NLO logarithmic scale dependence, the running of the coupling, without taking into account another source of logarithmic scale dependence, the renormalization of the $B$-field operators.
This operator renormalization also causes a logarithmically scale-dependent shift which is not captured by using an RG-flowed $g^2$ value in \cref{phiUV1}.
Therefore, in this work we will \textit{only} study the case where the UV is modeled by the NLO spectral function.

The NLO expression for the spectral function in the pure-glue theory is \cite{Banerjee:2022uge}
\begin{widetext}
    \begin{align}
        \phi^{\mathrm{NLO}}_{\mathrm{UV}}(\omega; \mubaromega) =
                                & c_B^2(\mubaromega, \mubarir)
        \left( \frac{g^2(\mubaromega) C_F \omega^3}{6\pi}
        \left\{ 1+\frac{N_c g^2(\mubaromega)}{(4\pi)^2}
        \left(\frac{5}{3}\ln\frac{\bar{\mu}_\omega^2}{4\omega^2}+\frac{134}{9}-\frac{2\pi^2}{3}\right) \right\}
        +\frac{g^4C_F}{12\pi^3}\left\{\cdots\right\}\right), \nonumber \\
        c_B^2 (\mu, \mubarir) = &
        \exp \left( \int_{\mubarir^2}^{\mu^2} \gamma_0  g^2(\mu)   \frac{\mathrm{d}\mu^2}{\mu^2} \right).
        \label{phiUV2}
    \end{align}
\end{widetext}
Note that, compared to \ccite{Banerjee:2022uge}, we have corrected \cite{delacruzinprep}
the term $-8\pi^2/3$ to $-2\pi^2/3$.
The expression $g^4 C_F/12\pi^3 (\ldots)$ in \cref{phiUV2}, which we have not completely written, is the thermal contribution; the remainder of the expression is the vacuum spectral function.
The expression for the thermal contribution is lengthy, but it turns out to be tiny compared to the vacuum contribution when $\omega \gg T$.%
\footnote{Unlike the stress tensor spectral function, the color-magnetic and color-electric spectral functions first deviate from their vacuum forms at NLO in the gauge coupling expansion.
    Structurally, this is because they involve the correlation of a single field strength with another, rather than a product of field strengths, so at the free level the correlation equals the sum of the vacuum-theory correlations summed over periodic copies. The result is that the thermal part first arises at NLO and is therefore small compared to the LO vacuum contribution.}
At very small frequencies $\omega < T$ it is not reliable and in any case small compared to the infrared contribution which the data require us to add.
We include the thermal part but have checked numerically that it has only a marginal effect on the spectral reconstruction.

\begin{figure*}
    \null \hfill
    \includegraphics[scale=1]{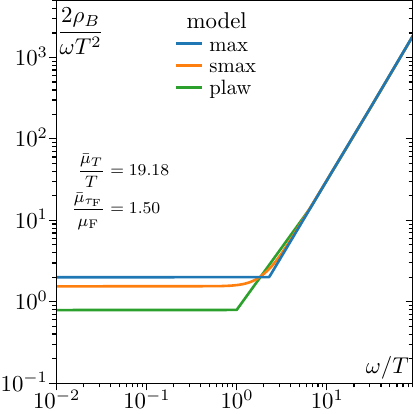}
    \hspace{2.1cm}
    \includegraphics[scale=1]{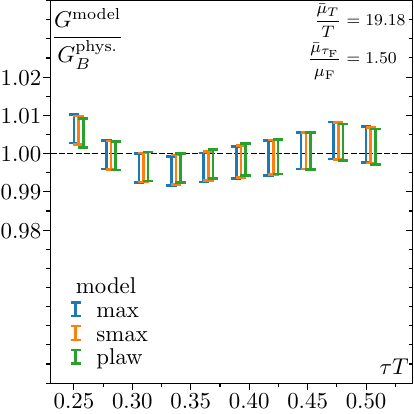}
    \hfill \null
    \caption{Left: fitted model spectral functions $\rho_B$ as a function of frequency $\omega$ in temperature units for the case $\mubarir=\mubarirNLO$, $\mubaruv=\mubaruvNLO$ (other choices differ only marginally). Statistical errors are not shown to reduce clutter; the intent of the figure is to show the differences in spectral function model shapes. The y-axis is scaled to yield $\kappa_B/T^3$ for $\omega\rightarrow 0$. Right: ratio of fitted model correlators $G^\mathrm{model}$ to renormalized, continuum, zero flow time correlator data $\GBphys$ (cf. \cref{fig:EEvsBB}) as a function of Euclidean time $\tau$ for the case $\mubarir=\mubarirNLO$, $\mubaruv=\mubaruvNLO$ (other choices differ only marginally). Different models are slightly offset in $\tau T$ for visibility.}
    \label{fig:spf-and-corr-fit-results}
\end{figure*}

The quantity in round parenthesis in \cref{phiUV2} represents the $\MSBAR$ spectral function.
This should be converted to the spectral function for the physical correlator to account for the renormalization of the magnetic field operators on the Wilson line.
This necessitates the factor $c_B^2$, which plays the same role as $\Zmatch$ played in \cref{sec:renorm} and in particular in \cref{phys-to-MS}.
Including this factor,
the expression in \cref{phiUV2} is formally independent of the renormalization scale $\mubaromega$ up to NNLO effects, since the explicit logarithm cancels with the implicit $\mu$ dependence from the one-loop running of $g^2(\mubaromega)$ and the $\mu$ dependence of $c_B^2$.
Nevertheless, we have to choose \textit{some} value, and we should attempt to pick a value which will somehow reflect the intrinsic scales present in the spectral function.
One such scale is the temperature, and the other is the frequency itself.
To estimate the typical thermal scale, we adopt the thermal scale of the dimensionally reduced 3D effective field theory \cite{Kajantie:1997tt},
\begin{align}
    \mudr=4 \pi  \exp \left(-\gamma_E-\frac{N_c-8 N_f \ln 2}{22 N_c-4 N_f}\right) T,
\end{align}
with $N_c=3$ and $N_f=0$ yielding $\mudr\approx 6.74T$.
This scale should be used for $\omega \sim T$.
For the relevant scale at large frequency we use $\mubar = \omega$, and we transition from one to the other via
\begin{align}
    \bar{\mu}_{\omega} \equiv \sqrt{\omega^2 + \mudr^2}.
\end{align}

To calculate $c_B^2$ we also again need the $\MSBAR$ coupling constant, which we reuse from \cref{sec:coupling}.
To take the systematics from higher-loop corrections into account we allow a rescaling of the UV part by the fit parameter $K$,
that is, we choose $\phi_{\mathrm{UV}} = K \phi^{\mathrm{NLO}}_{\mathrm{UV}}$.

The IR part manifests a very simple structure based on the infrared asymptotics~\cite{CaronHuot:2009uh},
\begin{align}
    \phi_{\rm{IR}}(\omega) \equiv \frac{\kappa_B \omega}{2T}.
    \label{phiIR}
\end{align}

Now that both IR and UV parts have been established, the missing link is the region between them.
Several schemes have been considered in the literature \cite{Francis:2015daa, Altenkort:2020fgs, Brambilla:2020siz, Banerjee:2022uge, Brambilla:2022xbd, Banerjee:2022gen, Altenkort:2023oms}.
In this study we employ three of them that we consider representative. The first one is simply to take the maximum of the two asymptotics \cite{Francis:2015daa}
\begin{align}
    \rho_{\rm{max}} \equiv \rm{max}(\phi_{\rm{IR}}, \phi_{\rm{UV}}).
    \label{model1}
\end{align}
The second one makes a smooth switch between the two parts \cite{Francis:2015daa}
\begin{align}
    \rho_{\rm{smax}} \equiv \sqrt{\phi_{\rm{IR}}^2+ \phi_{\rm{UV}}^2}.
    \label{model2}
\end{align}
The third one imposes a power-law transition between $\phi_{\rm{IR}}$ and $\phi_{\rm{UV}}$ \cite{Brambilla:2020siz}
\begin{align}
    \begin{split}
         & \rho_\mathrm{plaw}\equiv
        \begin{cases}
            \phi_\mathrm{IR} & \phantom{\mathrm{for}}\quad \omega \leq \omega_\mathrm{IR},         \\
            p(\omega)        & \mathrm{for}\quad \omega_\mathrm{IR} < \omega < \omega_\mathrm{UV}, \\
            \phi_\mathrm{UV} & \phantom{\mathrm{for}}\quad \omega \geq \omega_\mathrm{UV},         \\
        \end{cases}                                             \\
         & p(\omega) = a \omega^b,\quad a = \frac{\phi_\mathrm{IR}(\omega_\mathrm{IR})}{(\omega_\mathrm{IR})^b},                                      \\
         & b =  \frac{\ln\phi_\mathrm{IR}(\omega_\mathrm{IR})-\ln\phi_\mathrm{UV}(\omega_\mathrm{UV})}{\ln\omega_\mathrm{IR}- \ln\omega_\mathrm{UV}}.
    \end{split}
    \label{model3}
\end{align}
The choice for the cuts $\omega_\mathrm{IR}$ and $\omega_\mathrm{UV}$ can be justified based on some physical arguments. Common option of $\omega_\mathrm{IR}$ ranges from $T$ \cite{Kajantie:1997tt} to $\pi T$ \cite{Gubser:2006nz}. However, according to \ccite{Altenkort:2020fgs}, $\rho_{\rm{max}}$ and $\rho_{\rm{smax}}$ also tend to exhibit a transition around $\pi T$. This makes the choice $\pi T$ redundant so we use $\omega_\mathrm{IR}=T$.
As for $\omega_\mathrm{UV}$ we choose $2\pi T$, which is the standard thermal scale obtained from the perturbation theory in the $\overline{\mathrm{MS}}$ scheme \cite{Kajantie:1997tt}.

With these models we fit the lattice data by minimizing
\begin{align}
    \chi^2\equiv\sum_{\tau}\bigg{[}\frac{G_B(\tau)-G^\textrm{model}(\tau)}{\delta G_B(\tau)}\bigg{]}^2,
\end{align}
where $\delta G_B(\tau)$ denotes the error of the lattice data and $G^\textrm{model}(\tau)$ is calculated using the convolution (\cref{continuation}) replacing $\rho_B$ by the above models. The two free parameters of the spectral function models are $K$ and $\kappa_B/T^3$.

The results of the spectral function model fits are shown in \cref{fig:spf-and-corr-fit-results}. The left panel highlights the differences between the spectral function model shapes [defined in \cref{model1,model2,model3}] in the transition from infrared to ultraviolet asymptotics. The right panel of \cref{fig:spf-and-corr-fit-results} shows the residual deviation between the resulting fitted model correlator and the physical color-magnetic correlator data obtained in the previous section.
We remark that all the fit procedures in this study are performed on each of the 1000 bootstrap samples separately, with the individual fits using inverse squared statistical errors employed as weights.
All models describe the lattice data well according to the residual $\chi^2$/d.o.f., which is $1.8\pm0.8$ for $\rho_\mathrm{max}$, $1.6\pm0.8$ for $\rho_\mathrm{smax}$, and $1.2\pm0.7$ for $\rho_\mathrm{plaw}$. There is no meaningful difference in the fit quality between the different choices of $\mubarir$ and $\mubaruv$. The fit results for the $K$-factors are $K\approx 0.85\pm0.01$ for all models using $\mubarir=\mubarirLO$ and  $K\approx 1.03 \pm 0.01$ for all models using $\mubarir=\mubarirNLO$.

\begin{figure}[b]
    \centering
    \includegraphics[scale=1]{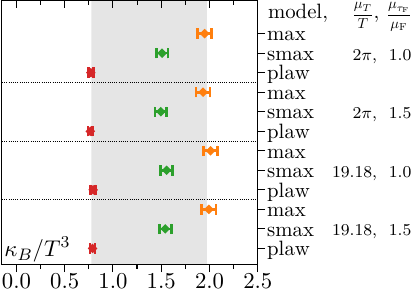}
    \caption{Coefficient $\kappa_B$ of the finite-mass correction to heavy-quark momentum diffusion from each model.
        The gray band shows the confidence interval that ranges from the 34th to 68th percentile of the total distribution of the fit results of all bootstrap samples for all models (see \cref{sec:extrap}).}
    \label{fig:scattering}
\end{figure}

By aggregating the fit results for $\kappa_B/T^3$ across different models, we obtain a scatter plot, as shown in \cref{fig:scattering}. The systematic and statistical errors, combined according to the methodology detailed in \cite{Altenkort:2023oms}, yield a confidence interval for $\kappa_B/T^3$, which is depicted as a gray band.
Including all four combinations for $\mubarir$ and $\mubaruv$ yields the final estimate
\begin{equation}
    0.78\leq \frac{\kappa_B}{T^3}\leq 1.97, \quad T=1.5T_c.
\end{equation}

\begin{figure}[t]
    \centering
    \includegraphics[scale=1]{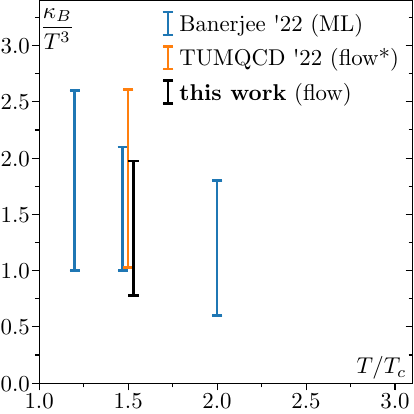}
    \caption{Comparison of quenched $\kappa_B/T^3$ obtained in this work (including the statistical and systematic uncertainty from all four choices of $\mubarir$ and $\mubaruv$, as well as the various spectral function models of \cref{sec:continue}) and recent literature. ``Banerjee '22 (ML)" refers to the multilevel method work of \ccite{Banerjee:2022uge}, while ``TUMQCD `22 (flow$^*$)" refers to \ccite{Brambilla:2022xbd}, in which gradient flow is used but the spectral reconstruction is performed at finite flow time, indicated by an asterisk in the legend.
    }
    \label{fig:compare-kappa}
\end{figure}

In \cref{fig:compare-kappa} we compare $\kappa_B$ obtained in this work (black line) and two other calculations, one using the multilevel method \cite{Banerjee:2022uge} and another also using gradient flow with analytical continuation performed at finite flow depth \cite{Brambilla:2022xbd}.
We find that the results are consistent even though they are calculated in different ways.

\begin{figure}[t]
    \centerline{
        \includegraphics{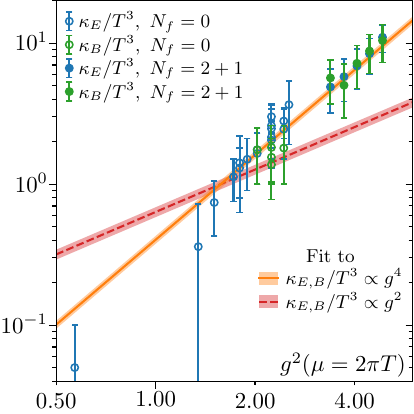}
    }
    \caption{
        Log-log plot of $\kappa_E/T^3$ and $\kappa_B/T^3$ versus the coupling strength $g^2(\mubar=2\pi T)$. The data, from
        \ccites{Francis:2015daa,Altenkort:2020fgs,Banerjee:2022gen,Brambilla:2022xbd,Brambilla:2020siz,Banerjee:2022uge} and this work,
        follow a single trend and turn out to fit well to a function proportional to $g^4$, given by $\kappa_{E,B}/T^3 = 0.4(0.03) g^4$ (orange curve, with the band being the error). The red curve shows an unsuccessful fit of form $ \kappa_{E,B}/T^3 \propto g^2$.
        The fit results are consistent with the perturbative picture that suggests $\kappa_{E,B}/T^3 \propto g^4$ \cite{Caron-Huot:2007rwy}.}
    \label{fig:kappa-g2}
\end{figure}

In \cref{fig:kappa-g2} we collect all existing determinations for $\kappa_E$ and $\kappa_B$ on the market and plot them logarithmically as a function of $g^2(\mu=2\pi T)$. The quenched data for $\kappa_E$ are taken from \ccites{Francis:2015daa,Altenkort:2020fgs,Banerjee:2022gen,Brambilla:2022xbd,Brambilla:2020siz}, while the quenched data for $\kappa_B$ is taken from \ccites{Brambilla:2022xbd, Banerjee:2022uge} and this work. The full QCD data for $\kappa_E$ and $\kappa_B$ are taken from \ccite{Altenkort:2023oms} and \ccite{Altenkort:2023eav}, respectively.
Based on perturbation theory \cite{Caron-Huot:2007rwy}, both $\kappa_E$ and $\kappa_B$ are expected to be proportional to $g^4$.
This is despite the UV behavior of the spectral function scaling as $g^2$, as shown in \cref{phiUV1}.
When we fit all $\kappa_B$ and $\kappa_E$ data using two different \textit{Ansätze}, one proportional to $g^2$ and the other proportional to $g^4$, we find that the former can describe neither $\kappa_E$ nor $\kappa_B$ and neither quenched nor unquenched data (red curve).
But a single fit of form $g^4$ actually describes all data (quenched and unquenched, $\kappa_E$ and $\kappa_B$) within errors (orange curve) with a residual $\chi^2/\mathrm{d.o.f.}=0.36$. This corroborates the validity of the perturbative prediction, which has also recently been used in an out-of-equilibrium study on heavy-quark diffusion \cite{Du:2023izb}.

\section{results and conclusions}
\label{sec:results}

In this paper we calculate the mass correction to the heavy-quark momentum-diffusion coefficient using quenched lattice simulations.
The correction is extracted from the color-magnetic field correlation functions measured under gradient flow.
In our opinion the most interesting part of this paper is \cref{sec:renorm}, in which we elaborate how to take care of the anomalous dimension existing for a magnetic field operator that is further complicated by the matching from the gradient flow scheme to its physical value.
By introducing a matching factor $\Zmatch$, we establish a multistep routine that bridges the gap between the object measured under gradient flow and its physical counterpart.
$\kappa_B$ obtained from the physical correlation functions in this work is consistent with previous quenched results calculated in different ways.
We also find that $\kappa_{E,B}$ calculated on the lattice, in both the quenched and the unquenched case, follows a form $\kappa/T^3 = 0.4g^4(\mubar=2\pi T)$ that shows the same scaling with the coupling strength as the perturbative calculation does \cite{Caron-Huot:2007rwy}.
This remarkably simple expression describes all quenched and unquenched lattice results with a good $\chi^2$.
The methodology developed in this work can be used in the future for a full QCD study at the physical pion mass reaching lower temperatures around the crossover region.

All data from our calculations, presented in the figures of this paper, can be found in Ref.~\cite{BBdatapublication}.

\section*{Acknowledgements}

This material is based upon work supported by the U.S. Department of Energy, Office of Science, Office of Nuclear Physics through Contract No.~DE-SC0012704, and within the frameworks of Scientific Discovery through Advanced Computing (SciDAC) award \textit{Fundamental Nuclear Physics at the Exascale and Beyond} and the Topical Collaboration in Nuclear Theory \textit{Heavy-Flavor Theory (HEFTY) for QCD Matter}.
The authors acknowledge support by the Deutsche For\-schungs\-ge\-mein\-schaft
(DFG, German Research Foundation) through the CRC-TR 211 'Strong-interaction matter under extreme conditions'– Project No. 315477589 – TRR 211. The computations in this work were performed on the GPU cluster at Bielefeld University.
We thank the Bielefeld HPC.NRW team for their support.

\end{document}